\pdfoutput=1
\documentclass{article}
\usepackage{spconf,amsmath,graphicx,multirow,tabularx,booktabs,array,hyperref}
\usepackage{cite}
\usepackage{bm}
\usepackage{stfloats}
\usepackage{graphics}
\usepackage{cleveref}

\def\z{{\mathbf z}}
\def\c{{\mathbf c}}
\def\q{{\mathbf q}}

\title{Language Adaptive Cross-lingual Speech Representation Learning with Sparse Sharing Sub-networks}
\name{Yizhou Lu, Mingkun Huang, Xinghua Qu, Pengfei Wei, Zejun Ma}
\address{
    Speech \& Audio Team, ByteDance AI Lab \\
    \small{\texttt{\{luyizhou.ritter, huangmingkun, xinghua.qu, pengfei.wei, mazejun\}@bytedance.com}}
}

\begin{document}
\ninept
\maketitle

\begin{abstract}
Unsupervised cross-lingual speech representation learning~(XLSR) has recently shown promising results in speech recognition by leveraging vast amounts of unlabeled data across multiple languages. However, standard XLSR model suffers from language interference problem due to the lack of language specific modeling ability. In this work, we investigate language adaptive training on XLSR models. More importantly, we propose a novel language adaptive pre-training approach based on sparse sharing sub-networks. It makes room for language specific modeling by pruning out unimportant parameters for each language, without requiring any manually designed language specific component. After pruning, each language only maintains a sparse sub-network, while the sub-networks are partially shared with each other. Experimental results on a downstream multilingual speech recognition task show that our proposed method significantly outperforms baseline XLSR models on both high resource and low resource languages. Besides, our proposed method consistently outperforms other adaptation methods and requires fewer parameters.
\end{abstract}
\begin{keywords}
representation learning, multilingual, language adaptation, speech recognition
\end{keywords}

\section{Introduction}
\label{sec:intro}
Automatic speech recognition (ASR) techniques have documented many success stories and been widely deployed in the wild~\cite{saon2017english,chiu2018state}. However, building a reasonable ASR system for practical application still requires tens of thousands of hours annotated data~\cite{chiu2018state}. It remains a challenging task to extend ASR system to around 7000 languages in the world as most languages are lack of training data. Self-supervised learning, where training targets are derived from the input itself, has recently shown promising results in speech recognition~\cite{oord2018representation, schneider2019wav2vec, chung2019unsupervised, liu2020mockingjay, ling2020deep, baevski2020wav2vec}. Recent work on self-supervised learning can be categorized into two main approaches: contrastive learning approaches that distinguish a true sample from negative ones in a latent space~\cite{oord2018representation, schneider2019wav2vec, baevski2020wav2vec}, and reconstruction-based approaches that directly predict future frames~\cite{chung2019unsupervised, ling2020deep} or reconstruct masked inputs~\cite{liu2020mockingjay, wang2020unsupervised}. Self-supervised learning provides an efficient way to learn from unlabeled data, and it allows the network to learn with fewer labeled data and generalize better. Its feasibility for monolingual speech recognition with limited amounts of labeled data has been shown in~\cite{baevski2020wav2vec}.

Self-supervised learning is further applied to multilingual setting in \cite{ conneau2020unsupervised}, namely cross-lingual speech representation learning~(XLSR). Multilingual pre-training simplifies the procedure of training individual seed models for each language by supporting multiple languages with a single model. Recent studies~\cite{conneau2020unsupervised, wang2021unispeech} also show that multilingual pre-training outperforms monolingual pre-training in low resource languages. From a multitask learning perspective, the common knowledge learned from each task shall help the other related tasks learn and generalize better~\cite{caruana1997multitask}. However, due to the vast diversity of pronunciation styles across different languages, the shared network often struggles in optimizing various languages simultaneously~\cite{lu2020bi, pratap2020massively, pham2021efficient}. While low resource languages benefit from joint training with similar languages, high resource languages often suffer from the negative transfer problem, resulting in inferior performance~\cite{conneau2020unsupervised}. Such performance degeneration becomes more significant as the model expands to more languages~\cite{pratap2020massively} or more training data~\cite{ li2021scaling}, which obstacles the application of pre-trained models on downstream multilingual tasks~\cite{li2020multilingual}.

In this work, we study language adaptive cross-lingual speech representation learning to alleviate the aforementioned interference problem. More importantly, we propose a novel language adaptive pre-training approach based on sparse sharing sub-networks~(S3Net). Sparse sharing architecture is initially proposed in~\cite{sun2020learning} to jointly learn sub-networks for multiple tasks and further applied in the field of neural machine translation~\cite{liang2021finding, lin2021learning, xie2021importance}. Inspired by that, we extract a sub-network for each language and all the sparsely shared sub-networks are jointly trained to learn language adaptive speech representations, with each language only updating its corresponding sub-network. The key idea of this work is that redundant parameters can be pruned out separately for each language with minor or no performance degradation, and these parameters discarded by one language can be further utilized by other languages to learn better representations. Besides, it automatically distributes both shared and language specific parameters at each layer, without requiring any additional language specific component~\cite{kannan2019large}. Given these designs, the learnt representations are expected to benefit more from positive transfer, while the negative transfer effects from dissimilar languages are mitigated. 

There are several ways of extracting sub-networks, and in this paper we mainly focus on two ideas: one follows the procedure of lottery ticket hypothesis~(LTH)~\cite{frankle2018lottery} and the other is based on taylor expansion~(TE)~\cite{molchanov2019importance}. We compare our S3Net with baseline XLSR model and several other adaptation methods~\cite{kannan2019large, gong2021layer}. Experimental results on a downstream multilingual task show the effectiveness of our proposed method. Specifically, S3Net yields an average 9.8\% and 7.4\% relative error reduction for XLSR base and large models respectively. Notably, it significantly improves the performance of high resource languages, achieving an average 17.8\% and 16.7\% relative error reduction. Moreover, S3Net also consistently outperforms other adaptation methods and requires fewer parameters.

\section{Related Work}
\label{sec:related_work}
Language interference problem has been investigated in many prior work on multilingual ASR. From a capacity perspective, \cite{conneau2020unsupervised} finds that enlarging model size alleviates the interference problem, and \cite{li2021scaling} scales up their model to 10 billion parameters to accommodate multiple languages. Another line of this research tends to retain the language specific modeling ability. It is observed that simply adding a one-hot language identity~(LID) vector to condition the multilingual model can boost the performance~\cite{toshniwal2018multilingual}. To better capture the language specific knowledge, previous studies usually augment networks with additional manually designed components, such as language specific weight matrices~\cite{pham2021efficient}, light weight adapters~\cite{kannan2019large}, decoupled multilingual encoders~\cite{lu2020bi} or decoders~\cite{pratap2020massively}. However, the inserted module size, structure and injection position are all important factors to consider, which requires additional efforts to fuse those additional modules into the original network~\cite{gong2021layer}.

We follow the line of language adaptation in this study. While previous work are mainly focused on supervised setting, to the best of our knowledge, this is the first work to apply language adaptive training to unsupervised pre-training models.

\section{Language adaptive pre-training with sparse sharing sub-networks}
\label{sec:lap}
\begin{figure}[t]
  \centering
  \centerline{\includegraphics[width=0.9\linewidth]{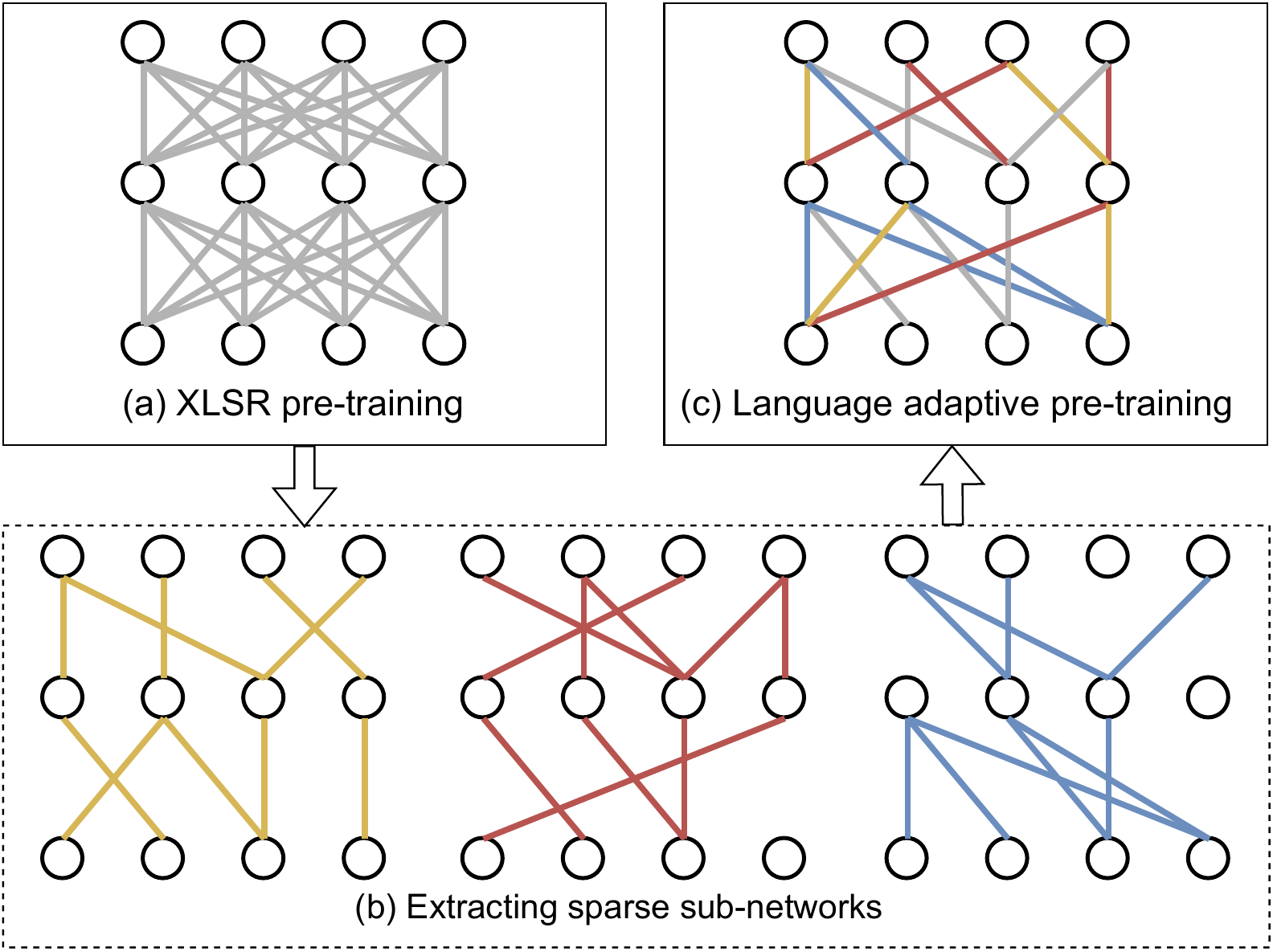}}
\caption{Training procedure of the proposed S3Net. Connections between two nodes can be shared by a set of languages (gray lines), or be occupied by one specific language (colored lines). There also exists empty connections in the final S3Net model.}
\label{fig:lap_procedure}
\end{figure}

In this section, we describe the proposed language adaptive pre-training approach based on sparse sharing sub-networks, namely S3Net. The proposed S3Net mainly includes three parts as shown in Figure~\ref{fig:lap_procedure}: XLSR pre-training, extracting sparse sub-networks and language adaptive pre-training.

\subsection{Pre-training of XLSR model}
\label{sec:xlsr}
XLSR model~\cite{conneau2020unsupervised} extends wav2vec 2.0 framework~\cite{baevski2020wav2vec} that basically consists of three components: feature encoder, context network and quantization module. 
The multi-layer convolutional feature encoder takes raw waveform as input and maps the input into latent speech representations $\z = \z_1, ..., \z_T$. Each $\z_t$ represents an approximately 25ms wide audio with the frame stride of 20ms. The transformer~\cite{vaswani2017attention} based context network utilizes the latent speech representations, and outputs contextual representations $\c=\c_1, ..., \c_T$ that capture contextual information from full sequence. To provide the training targets, the latent representation $\z_t$ is discretized to $\q_t$ with a quantization module. The quantizer has $G=2$ codebooks with $V=320$ entries each, resulting in a set of over 100K codewords in total. Gumbel softmax enables the model to choose discrete codebook entries in a fully differentiable way~\cite{jang2016categorical}.

The model is trained by solving a contrast learning task. Following~\cite{baevski2020wav2vec}, we randomly sample from all time steps with probability $s=0.065$ as the starting indices, and then continuously mask the subsequent $M=10$ steps. The goal of the contrast task is to distinguish a true encoded sample $\q_t$ among distractors $\mathbf{Q}_t$ that are sampled uniformly from other masked steps of the same sequence:
\begin{equation}
\label{eq:total_loss}
\begin{aligned}
  \mathcal{L} = -\log \frac{\exp (sim(\mathbf{c}_t, \mathbf{q}_t)/ \kappa)} {\sum_{\mathbf{\tilde{q} \in \mathbf{Q}_t}} \exp (sim(\mathbf{c}_t, \mathbf{\tilde{q}}) / \kappa)} + \lambda \mathcal{L}_d
\end{aligned}
\end{equation}
where $sim(\cdot,\cdot)$ denotes cosine similarity, $\lambda$ is the weight factor and diversity loss $\mathcal{L}_{d}=\frac{1}{G V} \sum_{g=1}^{G} \sum_{v=1}^{V} \bar{p}_{g, v} \log \bar{p}_{g, v} $ is employed to increase the use of codebook representations by maximizing the entropy over the selection probability of entry $v$ in codebook $g$.

We form multilingual batches~\cite{conneau2019cross2} by sampling with a multinomial distribution $p_l \sim (\frac{n_l}{N})^{\alpha}$, where $n_l$ is the number of hours for language $l$, $N$ is the total number of hours and $\alpha$ is the sampling factor. We upsample the data from low resource languages with $\alpha = 0.5$ and for high resource languages we use natural sampling probability with $\alpha = 1.0$.

\subsection{Extracting sparse sub-networks}
\label{sec:extract_mask}
Lottery ticket hypothesis~\cite{frankle2018lottery} suggests that, a randomly initialized dense network contains small sub-networks~(winning tickets) such that, when trained in isolation, can match the accuracy of full network. This motivates us to prune out redundant parameters for each language to make room for language specific modeling, and we hypothesize that these parameters discarded by one language can be further utilized by other languages to learn better representations. However, \cite{sun2020learning} shows that finding winning tickets from randomly initialized network is not stable, and a multitask warmup stage helps stabilize the process. Thus we extract sub-networks on top of a pre-trained XLSR model. Two different approaches of extracting sub-networks are explored in this work. The first one mainly follows the procedure of LTH~\cite{frankle2018lottery} and the second one is based on importance scores that are estimated with first order taylor expansion~\cite{molchanov2019importance}.

\textbf{Extracting sub-networks with LTH.}
For the first approach we adopt a simple one-shot magnitude pruning~(OMP) strategy instead of the iterative pruning strategy used in~\cite{frankle2018lottery}, as iterative pruning requires several rounds of training, pruning and resetting for each language, which is computationally expensive and time consuming. For each language $l$, we train XLSR model $\bm{\theta}$ with only the specific language data $\mathcal{D}^l$  for additional few steps to obtain $\hat{\bm{\theta}}^l$. Sub-network is extracted from $\hat{\bm{\theta}}^l$ and a proportion of $p$ parameters with lowest magnitude are considered as unimportant for this language, thus being pruned out. For the sake of convenience, we use a fix pruning rate $p$ for all languages in this work. The structure of sub-network for language $l$ is controlled by a binary mask matrix $\mathbf{m}^l$ which is initialized with $\mathbf{1}$. If a parameter $\hat{\theta}^l_i$ is pruned out, the corresponding item $m^l_i$ in the mask is updated to zero. The sub-network employed by language $l$ thus can be denoted as 
$
\bm{\theta}^l = \mathbf{m}^l \odot \bm{\theta}    
$,
where $\odot$ denotes the element-wise product operation.

\textbf{Extracting sub-networks with TE.}
Following~\cite{molchanov2019importance}, the importance of a parameter can be quantified by the error induced by removing it. A squared difference of prediction errors with and without the parameter $\theta_i$ is used to measure the importance score $\mathcal{I}^l_{i}$ of language $l$:
\begin{equation}
\label{eq:initial_importance_score}
  \mathcal{I}^l_{i}=\left[\mathcal{L}(\mathcal{D}^l, \bm{\theta})-\mathcal{L}\left(\mathcal{D}^l, \bm{\theta} \mid \theta_{i}=0\right)\right]^{2}
\end{equation}
where $\mathcal{D}^l$ denotes the training data of language $l$, and $\mathcal{L}(\mathcal{D}^l, \bm{\theta})$ is the averaged loss on $\mathcal{D}^l$ calculated by Equation~\ref{eq:total_loss}. However, directly calculating the importance score in Equation~\ref{eq:initial_importance_score} requires $|\bm{\theta}|$ times calculation of the loss function, which is infeasible. Fortunately, 
we can approximate the score with first order taylor expansion~\cite{molchanov2019importance}:
\begin{equation}
\mathcal{I}^l_{i} \approx (g^l_i \theta_i)^2
\end{equation}
where $g^l_i = \frac{\partial \mathcal{L}(\mathcal{D}^l, \bm{\theta}) }{\partial \theta_i}$ is the gradient for $\theta_i$ that can be efficiently calculated with backward propagation. For each language, those parameters with top $p$ lowest importance scores are pruned out, which is similar to the LTH based approach.

\subsection{Language adaptive pre-training with S3Net}
\label{sec:lap_with_s3net}
Once we obtain the masks $\mathbf{m}_1,...,\mathbf{m}_L$ for all languages, we continue to train the model $\bm{\theta}$ with all multilingual data to learn language adaptive speech representations. We form batches that only contain utterances from one language and those batches are randomly sampled with the same sampling strategy as XLSR pre-training. It is noted that each language only maintains a sub-network, and for each batch only the sub-network from the corresponding language will participate the forward computation and be updated. In contrast to other adaptation methods that typically require manually designed language specific components, S3Net automatically distributes both  shared and language specific parameters at each layer, and we expect that the model capacity is better allocated for each language.

\begin{table*}[t]
  \centering
\resizebox{2\columnwidth}{!}{
  \begin{tabular}{l|c|ccccccccc|c}
    \toprule
     {\textbf{Model}} & \textbf{Pre-trained data} & \textbf{es} & \textbf{fr} & \textbf{it} & \textbf{ky} & \textbf{nl} & \textbf{ru} & \textbf{sv} & \textbf{tt} & \textbf{zh} & \textbf{Avg} \\
    \midrule
    \multicolumn{2}{l|}{Number of unlabeled audio data} & 168h & 353h & 90h & 17h & 29h & 55h & 3h & 17h & 50h &  \\
    \midrule \midrule
    \multicolumn{11}{l}{{\it{Baselines from XLSR}}~\cite{conneau2020unsupervised}} \\
    \midrule
     {XLSR-Monolingual} & CV-Mono* & 6.8 & 10.4 & 10.9 & 29.6 & 37.4 & 11.6 & 63.6 & 21.4 & 31.4 & 24.8 \\
     {XLSR-10} & CV-Multi* &9.4 & 13.4 & 13.8 & 8.6 & 16.3 & 11.2 & 21.0 & 8.3 & 24.5 & 14.1 \\
     {XLSR-10 (Large)} & CV-Multi* & 7.7 & 12.2 & 11.6 & 7.0 & 13.8 & 9.3 & 20.8 & 7.3 & 22.3 & 12.4 \\
    \midrule 
    \multicolumn{11}{l}{\it{Re-run baselines and our models}} \\
    \midrule
    XLSR-10 & \multirow{3}{*}{CV-Multi} & 10.8 & 12.8 & 15.1 & 8.5 & 15.4 & 11.8 & 22.1 & 8.1 & 24.2 & 14.3 \\
    S3Net-TE & & 9.9 & 12.0 & 14.4 & 7.8 & 14.7 & 11.3 & 22.1 & 7.7 & 23.9 & 13.8 \\
    S3Net-LTH &  & \textbf{8.7} & \textbf{10.8} & \textbf{12.4} & \textbf{7.5} & \textbf{14.1} & \textbf{10.1} & \textbf{22.0} & \textbf{7.2} & \textbf{22.9} &  \textbf{12.9} \\
    \midrule
    XLSR-10 (Large) & \multirow{3}{*}{CV-Multi}  & 9.0 & 10.6 & 12.7 & 6.8 & 12.8 & 10.1 & 19.9 & 6.6 & 21.5 & 12.2 \\
    S3Net-TE (Large) & & 8.4 & 10.5 & 12.4 & 6.7 & 12.5 & 10.1 & 19.6 & 6.3 & 21.6 & 12.0 \\
    S3Net-LTH (Large) & & \textbf{7.3} & \textbf{9.2} & \textbf{10.4} & \textbf{6.3} & \textbf{12.1} & \textbf{9.4} & \textbf{19.5} & \textbf{6.1} & \textbf{21.5} & \textbf{11.3} \\
    \bottomrule
  \end{tabular}
}
  \caption{Evaluation results on CommonVoice dataset. The last column is the averaged PER on nine languages. Re-run baselines and our models are all pre-trained on ten languages, and evaluated on nine languages with shared vocabulary using CTC criterion. *: They use different version of the CommonVoice dataset, but the data size is the same as ours.}
  \label{tab:cv_all}
\end{table*}

\section{Experiments}
\label{sec:experiments}

\subsection{Experimental setup}
\label{subsec:exp_setup}

We use CommonVoice dataset~\footnote{\url{https://commonvoice.mozilla.org/datasets}. We use the December 2020 release version.}~\cite{ardila2019common} for pre-training. For a fair comparison with XLSR~\cite{conneau2020unsupervised}, we consider the following nine languages for evaluation~\footnote{There are initially ten languages for evaluation, but files of Turkish~(tr) in the test set are missing in CommonVoice December 2020 release, so we exclude this language.}: \textit{Spanish~(es), French~(fr), Italian~(it), Kyrgyz~(ky), Dutch~(nl), Russian~(ru), Swedish~(sv), Tatar~(tt) and Chinese~(zh).} We pre-train base and large models on 1350 hours unlabeled multilingual dataset~(CV-Multi) as in~\cite{conneau2020unsupervised}, which is composed of 782 hours data from above nine language plus additional 568 hours \textit{English~(en)} data. For fine-tuning, we use the evaluation splits from~\cite{riviere2020unsupervised}, which contains 1 hour labeled training data, 20 minutes validation data and 1 hour evaluation data for each language.

For our baseline models, we use the same model structure and training hyperparameters as XLSR~\cite{conneau2020unsupervised}. The models are all trained on 64 GPUs, and we train 250k steps for base model and 400k steps for large model. For fine-tuning, we adopt Connectionist Temporal Classification~(CTC)~\cite{graves2006connectionist} criterion and evaluate the multilingual performance of the pre-trained model. A randomly initialized output layer with shared vocabulary is added on top of the pre-training model. We use Adam optimizer and the learning rate is warmed up for 2k updates to 5e-5, keeps constant for 8k updates and then linearly decay for 10k updates. Phone error rate~(PER) is reported following previous work.

\subsection{Evaluation of proposed method}
\label{subsec:evaluation_s3net}

We denote S3Net using LTH for sub-network extraction as S3Net-LTH, and similarly S3Net-TE. For S3Net-LTH, we train the XLSR model with additional 50k steps separately for each language to extract the sub-networks. For S3Net-TE, we freeze the parameters and directly calculate importance scores with the gradients and weights. The prune rate is set to 0.4 for all base and large models, and by default the unstructured pruning is done layer by layer for each linear layer in context network. After obtaining the sub-network masks, we restart from the XLSR model and jointly train the sub-networks with all multilingual data for additional 50k steps. All hyper-parameters are tuned based on the performance of validation set. 

The results of baseline XLSR models and the proposed S3Net are shown in Table~\ref{tab:cv_all}. Specifically, we reproduce similar results as~\cite{conneau2020unsupervised}, with an average PER of 14.3\% and 12.2\% for base and large models. However, the results on each individual language are different from~\cite{conneau2020unsupervised} due to the different versions of pre-training data. We can see that both S3Net-TE and S3Net-LTH consistently outperform baseline XLSR models on each language, and S3Net-LTH models perform the best, with an average 9.8\% and 7.4\% relative PER reduction respectively for base and large models. Since high resource languages suffer more from language interference problem, it can be seen from the table that the proposed method achieves more performance improvements on high resource languages~(es, fr, it), with 17.8\% and 16.7\% average relative PER reduction. We also observe that S3Net-TE models perform worse than S3Net-LTH models. This is because S3Net-TE employs a simpler sub-network extraction strategy, which may not yield satisfactory sub-networks. Besides, we also conduct monolingual fine-tuning experiments individually on each languages, and we find that it performs slightly better than multilingual fine-tuning, with  average PER of 12.8\% and 11.2\% for S3Net-LTH base and large models.

\subsection{Comparison of S3Net with other adaptation methods}
\label{subsec:comparison_with_others}
We further compare the proposed method with other adaptation methods such as gating network~\cite{gong2021layer} and adapter~\cite{kannan2019large}. For gating network, one-hot LID embedding is used to learn the language specific scaling and biasing vectors~\cite{gong2021layer}. We add gating network module directly after the output of feature encoder to modulate the latent representations. As for the adapter based language adaptation method, we add adapters before the input of each transformer layer. The structure of adapter module follows~\cite{kannan2019large}, and the projection dimension is set to 256 for base and large model. The gating network and adapter modules are inserted into the XLSR model and further trained with 50k steps as S3Net models. We show in Table~\ref{tab:comp_adapt} that all the adaptation methods improve the performance of baseline XLSR model, and our proposed S3Net-LTH outperforms all other adaptation methods while requiring fewer parameters.
\begin{table}[htbp]
\vspace{-0.1in} 
\caption{Comparison of different adaptation methods. Multilingual evaluation results are averaged on high resource languages~(High), low resource languages~(Low) and all nine languages~(Avg).}
\label{tab:comp_adapt}
\centering
\resizebox{\columnwidth}{!}{
\begin{tabular}{l|c|ccc}
\toprule
\multicolumn{1}{l|}{\multirow{2}{*}{Model}} &
\multicolumn{1}{c|}{\multirow{2}{*}{\#Params}} &
\multicolumn{3}{c}{CV-Eval} \\ \cline{3-5} 
\multicolumn{1}{c|}{} & \multicolumn{1}{c|}{} & 
High            & Low             & Avg            \\ \midrule 
XLSR-10  & 95M & 12.9 & 15.0 & 14.3     \\
+ Gating Network & 95M & 12.2 & 14.7 & 13.9      \\
+ Adapter & 143M & 11.5 & 14.1 & 13.2 \\
S3Net-LTH & 95M & \textbf{10.6} & \textbf{14.0} & \textbf{12.9} \\
\midrule
XLSR-10 (Large)  & 317M & 10.8 & 13.0 & 12.2     \\
+ Gating Network & 317M & 10.4 & 12.8 & 12.0      \\
+ Adapter & 444M & 10.4 & 12.9 & 12.1 \\
S3Net-LTH (Large) & 317M & \textbf{9.0} & \textbf{12.5} & \textbf{11.3} \\
\bottomrule
\end{tabular}
}
\end{table}

\subsection{Ablation study of sparse sub-networks}
To analyze the influence of different sub-networks on the performance of S3Net, we perform extensive ablation studies in Table~\ref{tab:comp_sys}. In the standard setup, we individually extract sub-networks for each language~(\#Mask=10). To verify that the improvements of S3Net come from the language specific modeling, we jointly extract one sub-network for all languages (\#Mask=1). We also experiment with individually extracting sub-networks for high resource languages~(en, es, fr, it) but jointly extracting one sub-network for all low resource languages~(\#Mask=5). Besides, we find that layerwise pruning performs slightly better compared with global pruning. We also conduct randomly pruning experiments, which demonstrates the effectiveness of extracting sub-networks with proposed strategies.

\begin{table}[htbp]
\vspace{-0.1in} 
\caption{Analysis of different sub-networks. Models are trained with base structure and prune rate is set to 0.4 throughout the experiments.}
\label{tab:comp_sys}
\centering
\resizebox{\columnwidth}{!}{
    \begin{tabular}{l|ccc|ccc}
    \toprule
    \multicolumn{1}{l|}{\multirow{2}{*}{Model}} &
    \multicolumn{1}{c}{\multirow{2}{*}{\#Mask}} &
    \multicolumn{1}{c}{\multirow{2}{*}{Type}} &
    \multicolumn{1}{c|}{\multirow{2}{*}{Strategy}} &
    \multicolumn{3}{c}{CV-Eval} \\ \cline{5-7} 
    \multicolumn{1}{c|}{} & \multicolumn{1}{c}{} & \multicolumn{1}{c}{} & \multicolumn{1}{c|}{} &
    High            & Low             & Avg            \\ \midrule 
    XLSR-10 & N/A & N/A & N/A & 12.9 & 15.0 & 14.3 \\ \midrule 
    \multirow{6}{*}{S3Net} & 1  & Global & LTH & 13.0 & 15.3 & 14.5 \\
        & 5  & Global & LTH & 10.8 & 15.0 & 13.6 \\
        & 10 & Global & LTH & 10.8 & \textbf{14.0} & 13.0 \\
        & 10 & Global & Random & 14.2 & 16.9 & 16.0 \\
        & 10 & Layerwise & TE & 12.1 & 14.6 & 13.8 \\
        & 10 & Layerwise & LTH & \textbf{10.6} & \textbf{14.0} & \textbf{12.9} \\
    \bottomrule
    \end{tabular}
}
\end{table}

\subsection{Analysis of different prune rate}
\label{subsec:analysize_prune_rate}
We further investigate the influence of different prune rates. We gradually increase the prune rate $p$ from 0.0 to 0.8 for S3Net-LTH base model, and evaluate the multilingual performance on nine languages. Training XLSR model with additional 50k steps~($p=0.0$) slightly improved the PER from 14.3\% to 14.1\%. It is shown that the best choice of $p$ is 0.4 and the performance starts to degrade as the prune rate continues to increase. The figure also shows that we can prune out as many as 70\% parameters for each language, while the performance of S3Net-LTH still outperforms baseline XLSR model. Similar phenomenon is also observed in large models. 

\begin{figure}[!ht]
\vspace{-0.1in} 
\label{fig:prune_rate}
\centering
\includegraphics[width=0.8\linewidth]{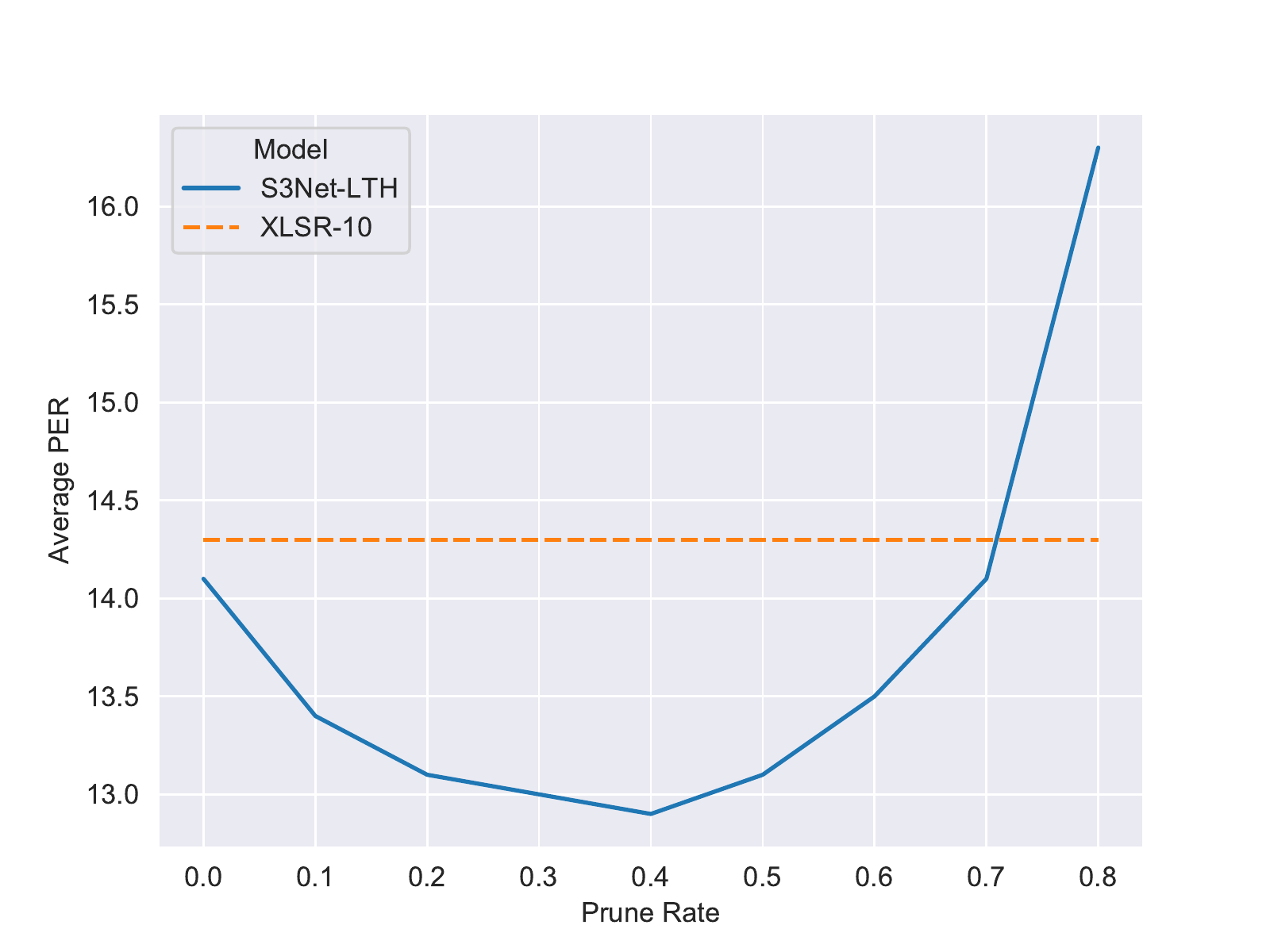}
\caption{Evaluation results of different prune rate for S3Net-LTH.}
\end{figure}

\section{Conclusion and Future Work}
\label{sec:future_work}

In this work, we study language adaptive cross-lingual speech representation learning. We investigate different approaches of extracting sub-networks and show that the proposed S3Net helps alleviating the language interference problem, especially for high resource languages. In the future, we plan to experiment with larger scale multilingual data and the application of multilingual pre-trained models on downstream tasks. We also plan to study structured sparsity and N:M sparsity for network acceleration.

\vfill\pagebreak

\bibliographystyle{IEEEbib}
\bibliography{refs}

\end{document}